\documentclass[9pt, conference]{IEEEtran}

\IEEEoverridecommandlockouts

\usepackage{cite}
\usepackage{amsmath,amssymb,amsfonts}
\usepackage{algorithmic}
\usepackage{graphicx}
\usepackage{tabularx}
\usepackage{textcomp}
\usepackage{xcolor}
\def\BibTeX{{\rm B\kern-.05em{\sc i\kern-.025em b}\kern-.08em
    T\kern-.1667em\lower.7ex\hbox{E}\kern-.125emX}}

\usepackage{hyperref}
\usepackage{graphicx}
\usepackage{url,times}
\usepackage{multirow}
\usepackage{siunitx}
\usepackage{pifont}
\usepackage{booktabs}
\usepackage{mathtools}
\newcommand{\norm}[1]{\left\lVert#1\right\rVert}

\begin{document}

\title{Code Drift: Towards Idempotent Neural Audio Codecs \\
\thanks{This work was supported by NSF Award Number 2222369.}
}

\author{

\IEEEauthorblockN{Patrick O'Reilly}
\IEEEauthorblockA{
\textit{Northwestern University}\\
}

\and
\IEEEauthorblockN{Prem Seetharaman}
\IEEEauthorblockA{\textit{Adobe Research} \\
}

\and
\IEEEauthorblockN{Jiaqu Su}
\IEEEauthorblockA{\textit{Adobe Research} \\
}

\and
\IEEEauthorblockN{Zeyu Jin}
\IEEEauthorblockA{\textit{Adobe Research} \\
}

\and
\IEEEauthorblockN{Bryan Pardo}
\IEEEauthorblockA{
\textit{Northwestern University}\\
}
}

\maketitle

\begin{abstract}
Neural codecs have demonstrated strong performance in high-fidelity compression of audio signals at low bitrates. The token-based representations produced by these codecs have proven particularly useful for generative modeling. While much research has focused on improvements in compression ratio and perceptual transparency, recent works have largely overlooked another desirable codec property -- \textit{idempotence}, the stability of compressed outputs under multiple rounds of encoding. We find that state-of-the-art neural codecs exhibit varied degrees of idempotence, with some degrading audio outputs significantly after as few as three encodings. We investigate possible causes of low idempotence and devise a method for improving idempotence through fine-tuning a codec model. We then examine the effect of idempotence on a simple conditional generative modeling task, and find that increased idempotence can be achieved without negatively impacting downstream modeling performance -- potentially extending the usefulness of neural codecs for practical file compression and iterative generative modeling workflows.

\end{abstract}


\section{Introduction}

Neural network-based codecs have shown promise in lossy compression of digital media across modalities such as video, images, and audio \cite{dcvc, deepimage, encodec}. Whereas signal-processing codecs rely on hand-crafted algorithms for identifying and removing redundant information within a signal, neural codecs learn to encode and decode signals from data directly.

For audio, neural codecs typically encode waveform signals as one or more sequences of discrete tokens, each drawn from a codebook. High compression ratios can be achieved with a combination of small codebooks (e.g. $2^{10}-2^{12}$ possible tokens) and low token sequence rates (e.g. $100 - 1000$ times less than the source audio). Because these token representations are far more compact than waveform audio, while retaining most perceptually relevant information, it is common to train generative models of audio directly on token sequences \cite{audiolm, musicgen}. Predicted token sequences can then be mapped to waveform audio via the decoder of the chosen codec.

\textit{Idempotence} -- the stability of a codec's decoded output, and thus its encoded representation, under multiple rounds of encoding and decoding -- is a desirable property for codecs. This is because codecs with low idempotence significantly degrade media under multiple encodings, and may therefore be poorly suited for real-world use. Such multiple-encoding scenarios can arise when digital media is repeatedly compressed for transmission, easier storage, and retargeting for different platforms. The increasing adoption of neural codec-based generative models for audio postprocessing and editing \cite{genhance, voicecraft, vampnet} may also result in repeated encoding and decoding of an audio file before a final version reaches listeners. While some signal-processing codecs explicitly address idempotence or ``tandem coding" \cite{jpeg, tandemcoding, tandem2, tandem3}, the idempotence of neural audio codecs has not been studied in depth. Rather, existing research has focused on \textit{perceptual transparency} -- the preservation of output quality under a single encoding.

Neural audio codecs are typically designed not only for data compression, but to facilitate generative modeling on encoded token representations. While there has been some study on the relationship between representation idempotence and generative models for image codecs \cite{idempotentperceptual}, we are aware of no such study for audio codecs.

\begin{figure}
\begin{center}
\includegraphics[width=0.49\textwidth]{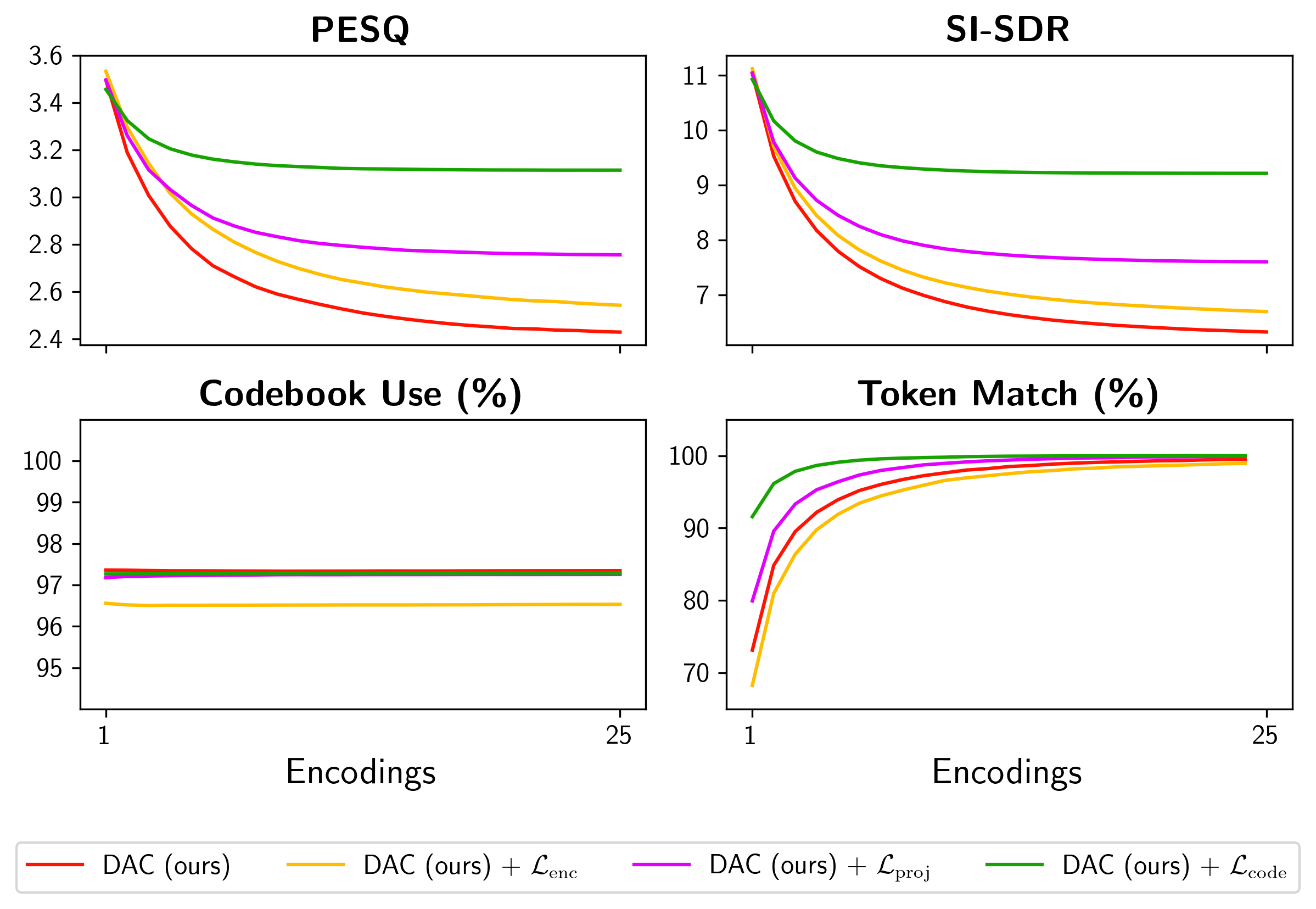}
\caption{\textbf{Effect of idempotent training}: audio quality (in PESQ and SI-SDR), average codebook use, and match rate between tokens encoded at the $n^{\mathrm{th}}$ and $(n + 1)^{\mathrm{th}}$ iterations as a function of encoding iterations for our DAC reproduction and fine-tuned variants with idempotence losses described in Section \ref{sec:encouraging}. Fine-tuning with $\mathcal{L}_{code}$ results in the highest overall idempotence through 25 encoding iterations, while preserving codebook use.} 
\label{fig:dac_variants}
\end{center}
\vspace{-1.0em}
\end{figure}

Motivated by the importance of idempotence and its potential impact on generative modeling of encoded representations, we undertake an empirical study of the idempotence of neural audio codecs. Our contributions are as follows:

\begin{itemize}
    \item We measure the idempotence of state-of-the-art neural audio codecs across speech datasets and find that many codecs demonstrate low stability under re-encoding.
    \item We develop a method for improving codec idempotence with minimal audio quality degradation through codec fine-tuning and evaluate the impact of different objective functions in encouraging idempotent representations.
  \item We explore the links between  idempotence and properties such as phase sensitivity, codebook use, and reconstruction quality.
    \item We evaluate the impact of codec idempotence on language modeling and find that idempotence can be increased without negatively affecting downstream modeling performance.
\end{itemize}

Through this work, we hope to encourage additional study of neural audio codec representations, and of their impact on both lossy compression and downstream generative modeling.
We provide audio examples at \url{https://oreillyp.github.io/codedrift/}.

\begin{table*}[h!]
\begin{center}
\caption{Comparison of Codecs}
\begin{tabular}{lcccccccc}
\toprule
\multirow{2}{*}{\textbf{Codec}} & \multirow{2}{*}{\textbf{Sample Rate (kHz)}} & \multirow{2}{*}{\textbf{Bitrate (kbps)}} & & \multicolumn{2}{c}{\textbf{VCTK}} & \multicolumn{2}{c}{\textbf{Expresso}} \\ 
\cmidrule(r){5-6} \cmidrule(r){7-8}
 &  &  & & \textbf{PESQ @ 1} $\uparrow$ & \textbf{PESQ @ 25} $\uparrow$ & \textbf{PESQ @ 1} $\uparrow$ & \textbf{PESQ @ 25} $\uparrow$ \\ 
\midrule
MP3 \cite{mp3} & 48 & 24 / 12 / 6 & & 3.79 / 3.79 / 3.41 & 1.99 / 1.99 / 2.29 & 3.49 / 3.49 / 3.32 & 1.93 / 1.93 / 2.13 \\
OPUS \cite{opus} & 48 & 24 / 12 / 6 & & 4.17 / 3.72 / 2.30 & 1.82 / 1.40 / 1.04 & 4.13 / 3.58 / 2.12 & 1.69 / 1.33 / 1.04 \\
\midrule
DAC \cite{dac} & 44.1 / 24 / 16 & 7.7 / 24 / 6 & & \textcolor{gray}{3.97 / 4.30 / 4.04} & \textcolor{gray}{3.17 / 1.28 / 1.32} & 3.78 / 4.29 / 3.91 & 2.89 / 1.24 / 1.27 \\
Encodec \cite{encodec, voicecraft} & 48 / 24 / 16 & 24 / 24 / 2.2 & & 2.96 / 3.65 / 3.03 & 1.05 / 1.05 / 2.21 & 2.87 / 3.34 / 2.91 & 1.05 / 1.05 / 2.08 \\
SpeechTokenizer \cite{speechtokenizer} & 16 & 4 & & 2.65 & 1.10 & 2.41 & 1.07 \\
FACodec \cite{naturalspeech3} & 16 & 4.8 & & 3.09 & 1.03 & 3.03 & 1.04 \\
Spectral-Codec \cite{spectralcodec} & 44.1 / 22.05 & 6.9 / 6.9 & & 3.24 / 3.49 & 1.18 / 1.36 & 3.16 / 3.41 & 1.14 / 1.27 \\
ESC \cite{esc} & 16 & 9 & & 3.85 & 2.50 & 3.67 & 2.24 \\
\bottomrule
\end{tabular}
\end{center}
\label{tab:comparisons}
\end{table*}

\section{Related Work}

\textbf{Neural Audio Codecs}: Recent works have proposed to use deep neural networks for audio compression \cite{soundstream, encodec, dac}. These neural codecs encode waveform audio into a hierarchy of token sequences through residual vector quantization (RVQ), and are trained end-to-end with objectives for both audio reconstruction quality and regularization of the encoded representations. Subsequent works have sought to disentangle encoded representations to correspond to ``semantic" attributes of audio such as prosody and linguistic content \cite{speechtokenizer, naturalspeech3}, and to augment encoded token sequences with time-invariant global information to better capture qualities such as speaker identity \cite{ticodec, naturalspeech3}. While the aforementioned codecs operate directly on waveform audio representations, recent works have found advantages in modeling magnitude and phase-magnitude spectrogram representations \cite{spectralcodec, esc}. In this work, we evaluate neural audio codecs from each of the above categories -- RVQ waveform codecs, disentangled ``semantic" codecs, and spectrogram codecs.

\textbf{Generative Modeling with Codec Representations}: While initial work on neural audio codecs focused on traditional lossy compression applications such as music and video streaming \cite{soundstream, encodec}, researchers have found that the encoded representations produced by these systems facilitate generative modeling of audio for tasks such as music generation \cite{musicgen, vampnet}, text-to-speech synthesis \cite{soundstorm, speechtokenizer}, sound effect generation \cite{audiogen}, and speech enhancement \cite{genhance}. These generative models are trained to predict token series using a language-modeling (categorical cross-entropy) objective in token space. In this work, we evaluate the performance of neural audio codecs on a simple conditional generation task (vocoding) in which tokens are predicted from a heavily compressed mel-frequency cepstrum coefficient (MFCC) representation.


\textbf{Studies of Neural Audio Codec Representations}:
Recent works have evaluated the utility of neural audio codec representations for both discriminative and generative downstream tasks \cite{dasb, codecsuperb, selfservingreviewer1, selfservingreviewer2}. Others have focused on identifying specific properties within these representations and understanding their impact on downstream performance. Lemercier et al. \cite{independenceloss} find that the 
codebooks learned by Encodec \cite{encodec} exhibit conditional dependence and show that downstream generative modeling performance can be improved by reducing this dependence via an additional training objective. Similarly, we evaluate one specific codec property (idempotence) and examine its relationship to downstream generative modeling performance. 

\textbf{Idempotence in Neural Image Codecs}: Neural codec idempotence has been studied in the image modality. Zhang et al. \cite{evaluatingstrongidempotence} introduce the notion of ``strong idempotence" -- stability to re-encoding across different bandwidths -- and evaluate both traditional and neural image codec models. Xu et al. \cite{idempotentperceptual} explore the relationship between perceptual image compression and codec idempotence, and show that high-quality compression can be achieved by pairing a neural codec with an unconditional generative model through an ``inversion" optimization. Li et al. \cite{idempotentrightinverse} 
relax invertibility constraints on normalizing flow-based image compression to obtain a fully idempotent codec and improve the trade-off between compression ratio and perceptual transparency in near-idempotent variants. Kim et al. \cite{instabilitysuccessive} evaluate the idempotence of neural image codecs and propose an auxiliary training objective to improve idempotence; we evaluate similar objectives in our experiments. In this work, we extend the study of neural codec idempotence to the audio modality.

\section{Measuring Idempotence}
\label{sec:measuring}

We begin by evaluating the idempotence of state-of-the-art neural audio codecs in terms of the stability of both audio outputs and encoded representations under multiple rounds of encoding.

\textbf{Codecs}: We consider the waveform RVQ codecs Encodec (in the original configurations, as well as the 16kHz speech-domain variant trained by Peng et al.) \cite{encodec, voicecraft} and Descript Audio Codec (DAC) \cite{dac}; the ``semantic" codecs SpeechTokenizer  \cite{speechtokenizer} and FACodec \cite{naturalspeech3}; and the spectrogram-based spectral codecs of Langman et al. \cite{spectralcodec} and ESC \cite{esc}. We use the available pre-trained weights for all codecs, and evaluate each codec at the maximum supported bitrate. We note that these codecs differ substantially in native sample rate, compression ratio, training data, and architecture, making exact performance comparisons difficult; nevertheless, we find strong shared trends in the idempotence of audio outputs and encoded representations. Finally, as baselines we evaluate the signal-processing MP3 \cite{mp3} and OPUS \cite{opus} codecs at low bitrates (see Table \ref{tab:comparisons}).\\
\indent \textbf{Datasets} While many codecs are trained for both speech and general-purpose audio compression \cite{encodec, dac}, we focus on speech data to facilitate comparisons. We evaluate codecs on the full-bandwidth speech datasets Expresso \cite{expresso}, containing spontaneous and read speech across four speakers; and VCTK \cite{vctk}, containing read speech across 109 speakers.
For each dataset, we select 500 one-second excerpts for evaluation and resample audio to the native sample rate of each codec. For the signal-processing baselines, we select 100 excerpts, due to longer encoding times.
\\
\indent \textbf{Metrics}: To measure the stability of output audio under multiple rounds of encoding, we use the full-reference metrics \textit{Scale-Invariant Signal-to-Distortion ratio} (\textit{SI-SDR}) \cite{sisdr} and \textit{Perceptual Evaluation of Speech Quality} (\textit{PESQ}) \cite{PESQ}. PESQ estimates ``perceptually-weighted" similarity between the original audio and encoded/decoded audio (up to 8kHz), while SI-SDR operates on the available bandwidth. Given the original audio input as a reference, codecs with high idempotence should produce reconstructions with high PESQ and SI-SDR scores even after multiple encoding iterations. To analyze drift in the internal representations of neural codecs after multiple encoding iterations, we compute the \textit{match rate} between tokens at each separate codebook level of the encoded representation after $n$ and $(n + 1)$ encoding iterations. This value ranges from $100\%$ when all tokens match to $0\%$ when none do. While this is a coarse measure that does not account for codebook size or latent similarity, it allows us to observe the convergence of certain codecs towards ``stable tokenizations" over multiple iterations. We also measure \textit{codebook use} (the entropy of encoded token distributions as a percentage of the maximum achievable codebook entropy) to study whether codecs converge to small subsets of tokens after multiple encoding iterations. Higher entropy indicates broader codebook use. 

\begin{figure}
\includegraphics[width=.49\textwidth]{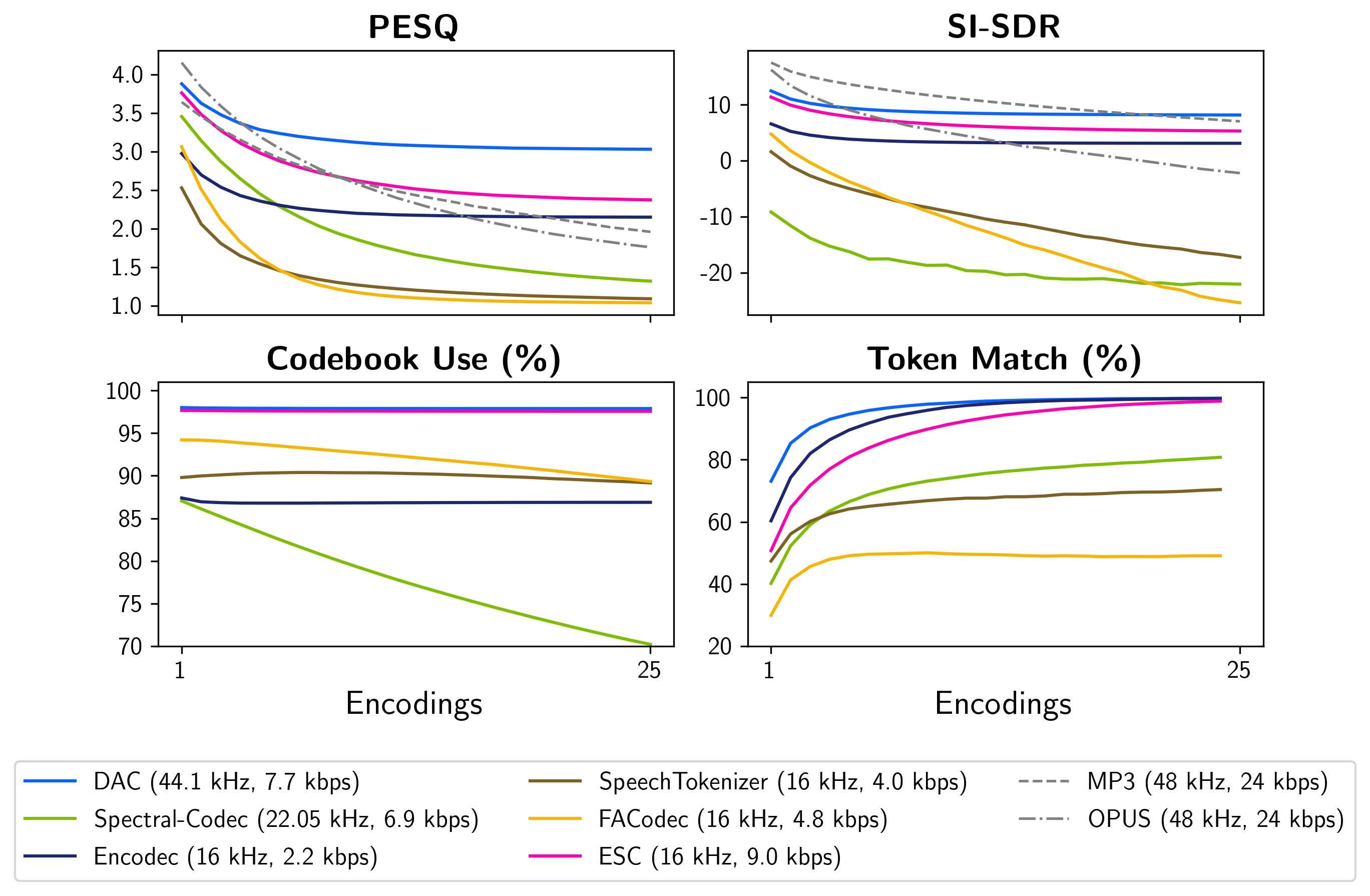}
\caption{\textbf{Audio and code stability over encodings}: 
audio quality (in PESQ and SI-SDR) of selected neural codecs by encoding iteration, average codebook use by encoding iteration, and match rate between tokens encoded at the $n^{\mathrm{th}}$ and $(n + 1)^{\mathrm{th}}$ iterations.} 
\label{fig:quality_and_codebooks}
\end{figure}
\begin{figure}
\begin{center}
\includegraphics[width=.49\textwidth]{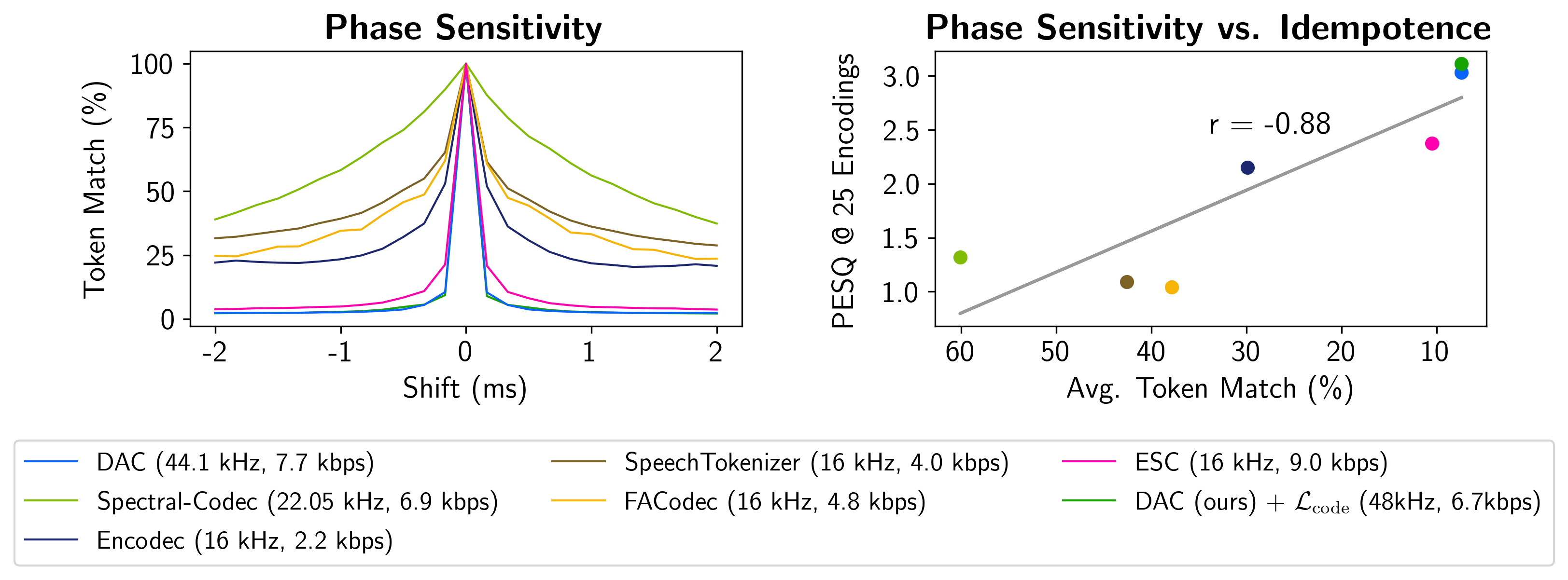}
\caption{\textbf{Phase sensitivity correlates with idempotence}: For each selected codec, we compute the token match rate between encodings of audio inputs under  time shifts between $\pm 2$ms. We then compute the correlation between average observed match rate and PESQ score after 25 encoding iterations. In general, codecs with higher phase sensitivity (as indicated by lower match rates) exhibit higher idempotence (as indicated by higher PESQ scores).
} 
\label{fig:robustness}
\end{center}
\vspace{-1.0em}
\end{figure}

In Table \ref{tab:comparisons}, we compare the performance of our selected codecs in terms of PESQ scores after a single encoding (indicating perceptual transparency) and after 25 encodings (indicating idempotence). Values are grey for datasets seen during codec training. Where we evaluate multiple variants of a codec (e.g. at different bitrates) we separate entries with slashes. In all experiments we normalize codec outputs to match the input volume. In Figure \ref{fig:quality_and_codebooks} we select the variant of each neural codec with the highest demonstrated idempotence and plot audio quality, codebook use, and token match rate as a function of the number of encoding iterations. 
In Figure \ref{fig:robustness} we plot the token match rate between encodings of an input under small time shifts, indicating the sensitivity of the token representation to phase, and compute the correlation between average observed sensitivity and idempotence as measured by PESQ score after 25 encoding iterations. 

The results of our experiments show that the idempotence of neural audio codecs varies greatly; DAC, ESC, and the Peng et al. variant of Encodec fare best, with DAC in particular demonstrating high audio quality even after 25 encoding iterations. Codec type alone (RVQ, semantic, or spectrogram) does not dictate performance. As can be seen in Figure \ref{fig:quality_and_codebooks}, the behavior of token representations under multiple encodings also varies significantly across codecs. In codecs exhibiting the highest idempotence (DAC, ESC), codebook use remains virtually unchanged, while token match rate increases -- indicating that even as these codecs converge to nearly fixed tokenizations, they do not collapse to a small subset of possible tokens. By contrast, Spectral-Codec and FACodec converge more slowly or not at all, while codebook use declines noticeably.

The observed differences in idempotence are not completely explained by the single-encoding perceptual transparency of codecs, as indicated by both Figure \ref{fig:quality_and_codebooks} and Table \ref{tab:comparisons}. For instance, DAC 24kHz and DAC 16kHz have high perceptual transparency (as indicated by PESQ and SI-SDR scores after one encoding iteration) but exhibit very low idempotence after 25 iterations. Phase sensitivity seems to play a larger role: as shown in Figure \ref{fig:robustness}, the sensitivity of token representations to phase shift positively correlates with idempotence. This is counterintuitive, as we might expect more idempotent codecs to be invariant to imperceptible perturbations such as phase shift. However, from our experiments it appears that fine-grained encoding of phase information is linked to the preservation of audio quality over multiple encoding iterations.

\section{Encouraging Idempotence}
\label{sec:encouraging}

Given the wide range of idempotence observed across neural audio codecs, we explore whether it is possible to improve idempotence without sacrificing perceptual transparency or fundamentally altering the compression scheme. We focus our efforts on DAC, as it shows the highest overall idempotence of audio and encoded representations in our experiments.

For a neural codec $f$ and input signal $x$, idempotence can be expressed as the relationship
$$
f(x) = f(f(x))
$$

\noindent A neural codec $f$ typically consists of an encoder network $f_e$, decoder network $f_d$, and quantization module $f_q$ such that its output can be expressed as $f(x) = f_d \circ f_q \circ f_e(x)$, where $f_q(x)$ is the quantized encoded representation. For DAC codec $f$ and input $x$, we can decompose the compression and decompression process into individual steps. Taking $ x' \coloneq f(x)$, we have:
\begin{equation}
    x' =  f_d \circ f_{dp} \circ f_{dq} \circ f_q \circ f_p \circ f_e(x) \\
\end{equation}
\noindent where $f_e$ is a neural network encoding waveform audio $x$ into frame-level continuous latents; $f_p$ is a linear projection into a low-dimension space for codebook lookup; $f_q$ is a $L_2$-normalized nearest-neighbors lookup over codebook vectors, returning codebook indices; $f_{dq}$ indexes the codebook and to obtain codebook vectors; $f_{dp}$ is a linear projection of the indexed codebook vectors back to the latent space; and $f_d$ is a neural network that upsamples the resulting latent sequence to obtain waveform audio. Because the vector quantization operation $f_{dq} \circ f_q$ is non-differentiable with respect to the input, the straight-through gradient estimator is used during backpropagation. DAC is trained with a combination of adversarial and spectrogram losses to encourage perceptual transparency as well as the standard codebook and commitment losses \cite{vqvae}. For additional details we refer the reader to Kumar et al. \cite{dac}. 

Similar to Kim et al. \cite{instabilitysuccessive}, we encourage idempotent representations through the use of an additional regularizing objective during codec training. We explore three regularizing losses for enforcing idempotence at different stages of the compression process.
\begin{enumerate}
    \item \textbf{Encoder idempotence}, the distance between continuous latents produced by the first and second encoding of a signal: \\
    \centerline{$\mathcal{L}_{\mathrm{enc}} = \norm{f_e(x') - f_e(x)}_2$}
    \item \textbf{Projected idempotence}, the distance between projected latents produced by the first and second encoding of a signal: \\ \centerline{$\mathcal{L}_{\mathrm{proj}} = \norm{f_p \circ f_e(x') - f_p \circ f_e(x)}_2$}
    \item \textbf{Token idempotence}, the distance between projected latents from the second encoding and the codebook vectors selected in the first encoding: \\ 
    \centerline{$\mathcal{L}_{\mathrm{code}} = \norm{f_p \circ f_e(x') - f_{dq} \circ f_{q} \circ f_p \circ f_e(x)}_2$}
\end{enumerate}

\noindent All losses are averaged over the time dimension of the representation at which they are computed. We note that the encoder idempotence loss is most similar to the idempotence loss proposed by Kim et al. \cite{instabilitysuccessive}, which is computed on the quantized latents of the first encoding; we find this degrades audio quality significantly and thus use the un-quantized latents instead. 

To evaluate the effect of our proposed idempotence losses, we train a reproduction of DAC from scratch for 500k steps on a mixture of openly-licensed speech data (CommonVoice \cite{commonvoice}, LibriVox \cite{librivox}, VoxPopuli \cite{voxpopuli}) and proprietary licensed sound effects and music data. Our DAC operates at 48kHz with a hop length of 640 samples, resulting in a token rate of 75Hz and bitrate of 6.7kbps; all other aspects of the model and training are identical to that of Kumar et al. After pretraining we freeze the quantizer module, disable codebook dropout, introduce one of the idempotence losses described above weighted by a constant $\lambda$, and fine-tune for a further 50k steps. To select $\lambda$ we sweep candidate values of $\{1, 10, 100, 1000\}$ and select the largest value for which the pretrained model's audio quality is preserved (as measured by validation loss). We use $\lambda = 1$ for $\mathcal{L}_{enc}$, $\lambda = 10$ for $\mathcal{L}_{proj}$, and $\lambda = 100$ for $\mathcal{L}_{code}$. In Figure \ref{fig:dac_variants} we perform the multiple-encoding experiment from Section \ref{sec:measuring} with our fine-tuned DAC variants and find that the choice of loss has a significant effect on the stability of output audio and encoded representations. All idempotent fine-tuned variants exhibit improved stability in audio quality under re-encoding when compared to our DAC reproduction. The codebook-space objective $\mathcal{L}_{\mathrm{code}}$ performs best, resulting in superior audio quality to all evaluated codecs (including the original DAC model of Kumar et al.) after 25 encoding iterations. The projected objective $\mathcal{L}_{\mathrm{proj}}$ is less effective, followed by the latent objective $\mathcal{L}_{\mathrm{enc}}$.

\section{Downstream Generative Modeling}

While idempotence is in itself valuable for applications such as file compression, many neural audio codecs are explicitly designed to facilitate generative modeling on encoded representations. Accordingly, we evaluate the impact of our idempotent training configurations on a simple generative modeling task. Using token representations obtained from DAC, our DAC reproduction, and our fine-tuned variants, we train SoundStorm-like \cite{soundstorm} masked generative transformer models to generate speech conditioned on 30-bin MFCC representations computed at the codec frame rate. This can be thought of as a vocoding task, albeit conditioned on a more compressed speech representation than the typical 80-120 band mel-spectrogram features \cite{cargan, bigvgan}, making token prediction difficult. We apply a proprietary bandwidth extension algorithm to the LJSpeech dataset \cite{ljspeech} to obtain roughly 24 hours of high-quality single-speaker read speech at 48kHz, allowing us to fully utilize the bandwidth of our DAC-based codecs. We train all models for 100k iterations on 5-second random excerpts at a batch size of 32 using the AdamW \cite{AdamW} optimizer with learning rate $1e-4$, and perform conditioning dropout with probability 0.2. We use small 43m-parameter transformer models consisting of 12 standard transformer encoder blocks with 8 attention heads each and apply rotary positional encodings in each block. During inference, we proceed from the lowest to highest codebook level following the ``MaskGIT" strategy used by Yang et al. \cite{genhance}; we use the confidence-based masking scheme from Garcia et al. \cite{vampnet} with mask temperature 10.5, sample with temperature 1.0 and nucleus 0.85 \cite{nucleus}, and apply classifier-free guidance with weight 2.0 \cite{classifierfreeguidance}. For each model we generate 100 samples using MFCC feature sequences from unseen recordings and compute the following metrics: \textit{SQUIM Mean Opinion Score (SQUIM-MOS)} \cite{squim}, a deep neural network-based estimate of speech quality that does not require precise alignment to the reference audio; 
\textit{Character Error Rate (ASR-CER)} of a Wav2Vec2.0 \cite{wav2vec2} speech recognition system's predicted transcripts on generated audio, which measures the intelligibility of generated speech; and PESQ. 
The results of our evaluations are presented in Table \ref{tab:generative}. Due to the information bottleneck of the MFCC representation, generations do not match the precise spectral structure of the source audio, as indicated by low PESQ scores. However, we find the generations retain high audio quality and intelligibility, as indicated by SQUIM-MOS and ASR-CER scores. Generative models trained on codecs fine-tuned with idempotence objectives produce outputs of similar quality to a model trained on the base codec, indicating that the proposed idempotence losses do not have a strong negative effect on generative modeling.

\begin{table}
\begin{center}
\caption{Generative Modeling on Codec Tokens}
\begin{tabular}{lcccc}
\toprule
\textbf{Codec} & & \textbf{SQUIM-MOS} $\uparrow$ & \textbf{PESQ} $\uparrow$ & \textbf{ASR-CER} $\downarrow$\\ 
\midrule
DAC (ours) & & 4.39 & 1.68 & 0.07 \\
DAC (ours) + $\mathcal{L}_{\mathrm{enc}}$ & & 4.34 & 1.66 & 0.06 \\
DAC (ours) + $\mathcal{L}_{\mathrm{proj}}$ & & 4.36 & 1.67 & 0.07 \\
DAC (ours) + $\mathcal{L}_{\mathrm{code}}$ & & 4.26 & 1.65 & 0.06\\
\bottomrule
\end{tabular}
\end{center}
\label{tab:generative}
\vspace{-1.0em}
\end{table}

\section{Conclusion}
\label{sec:discussion}

In this work we evaluate the idempotence of neural audio codecs and highlight possible connections to perceptual transparency, codebook use, and phase sensitivity. We then demonstrate that neural codec idempotence can be significantly increased via fine-tuning without sacrificing perceptual transparency or suitability for downstream generative modeling. This suggests that the idempotence of neural audio codecs can be improved without compromising desirable distributional properties of their encoded representations, allowing practical use in low-bitrate lossy file-compression scenarios without significant architectural or training modifications. Moreover, as codec-based generative models see wider use for audio editing tasks \cite{genhance} and in iterative creative interfaces \cite{tokentelephone}, idempotent fine-tuning may help to prevent the deterioration of audio quality under repeated applications of these models. We leave the study of the effects of codec idempotence on specific generative applications to future work. Future work may also explore incorporating idempotence objectives into codec training more fully; examining the impact of training data and architecture on idempotence; and applying idempotent codec architectures from the image modality such as invertible networks \cite{idempotentrightinverse} to naturally invertible audio representations (e.g. STFT).

\pagebreak

\bibliographystyle{IEEEtran}
\bibliography{refs}
\end{document}